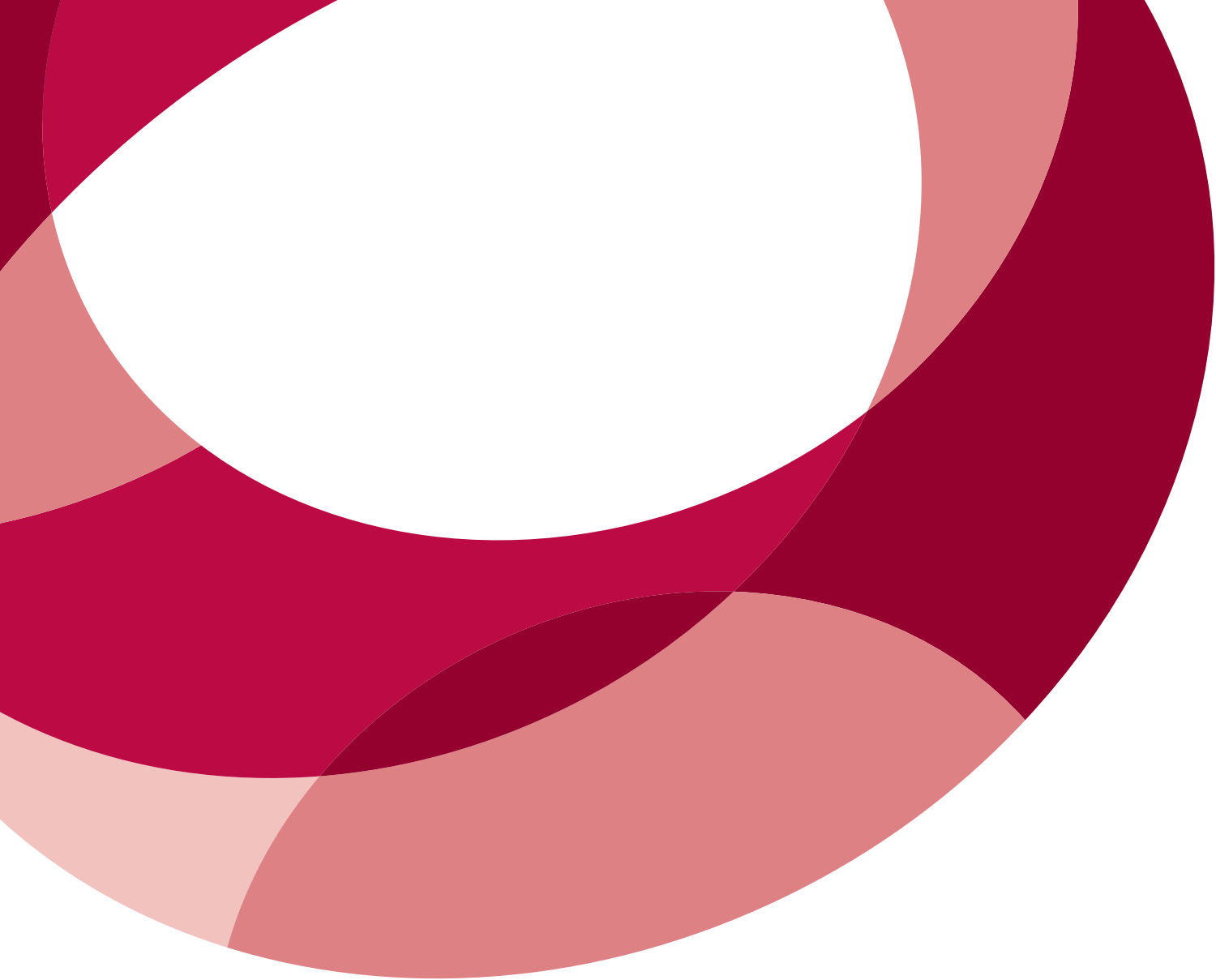

# Computational Support for Substance Use Disorder Prevention, Detection, Treatment, and Recovery

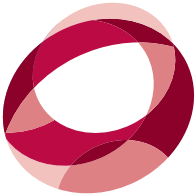

CCC
Computing Community Consortium
Catalyst

The material is based upon work supported by the National Science Foundation under Grants No. 1136993 and No. 1734706. Any opinions, findings, and conclusions or recommendations expressed in this material are those of the authors and do not necessarily reflect the views of the National Science Foundation.

# Computational Support for Substance Use Disorder Prevention, Detection, Treatment, and Recovery

**Workshop Chair**

Lana Yarosh, University of Minnesota

**Steering Committee**

Suzanne Bakken (Columbia University); Alan Borning (University of Washington); Munmun De Choudhury (Georgia Institute of Technology); Cliff Lampe (University of Michigan); Elizabeth Mynatt (Georgia Tech); Stephen Schueller (University of California Irvine); Tiffany Veinot (University of Michigan)

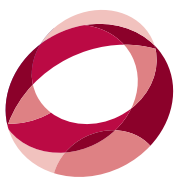

CCC
Computing Community Consortium
Catalyst

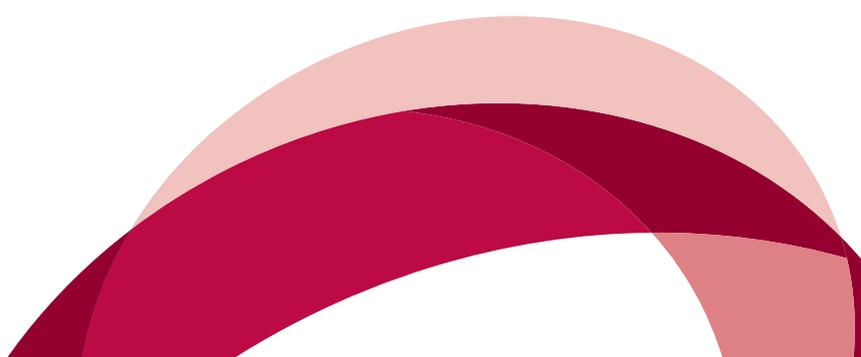

# 1. Overview

Substance Use Disorders (SUDs) involve the misuse of any or several of a wide array of substances, such as alcohol, opioids, marijuana, and methamphetamine. SUDs are characterized by an inability to decrease use despite severe social, economic, and health-related consequences to the individual. A 2017 national survey identified that 1 in 12 US adults have or have had a substance use disorder [5]. The National Institute on Drug Abuse estimates that SUDs relating to alcohol, prescription opioids, and illicit drug use cost the United States over $520 billion annually due to crime, lost work productivity, and health care expenses [2]. Most recently, the US Department of Health and Human Services has declared the national opioid crisis a public health emergency to address the growing number of opioid overdose deaths in the United States [20]. In this interdisciplinary workshop, we explored how computational support — digital systems, algorithms, and sociotechnical approaches (which consider how technology and people interact as complex systems) — may enhance and enable innovative interventions for prevention, detection, treatment, and long-term recovery from SUDs.

> **Substance Use Disorders (SUDs).** In the context of SUDs, a "substance" is any psychoactive compound, legal or illegal, which has "the potential to cause health and social problems" [34]. Use or misuse of substances may lead to an SUD, which is a diagnosable illness characterized by increasing tolerance, cravings, withdrawal symptoms, and inability to stop or reduce use despite increasing social, financial, and health consequences. A severe or chronic Alcohol Use Disorder may be colloquially referred to as "alcoholism," while chronic disorders pertaining to other substances are frequently referred to as "addiction."

Participants in this workshop approached computational support for SUD disorders through a variety of disciplinary lenses and approaches, including:

- **Human-Computer Interaction (HCI)** – HCI researchers develop new interactive systems (e.g., mobile health applications, virtual reality interventions, computer-mediated communication technologies), empirically investigate existing systems, and develop new approaches to manage information, augment community and peer support, and foster behavior change. HCI practitioners (also known as UI or UX professionals) design, implement, and evaluate the user experience of interactive technologies. SUDs constitute a compelling context in which to apply these approaches.

- **Ubiquitous Computing (Ubicomp)** – Ubicomp researchers innovate novel sensors and mobile computing analysis techniques to support gathering and acting on information about human activity in the world. Both positive and risky SUD behaviors may be detected and understood with Ubicomp techniques.

- **Informatics & Data Science** – Informatics and Data Science researchers leverage and innovate statistical and computational analysis techniques applied to large data sets (e.g. electronic health records (EHRs), social media traces, etc.) to advance scientific, community, or individual understanding of SUD-related behaviors and outcomes.

- **Behavioral and Mental Health** – Behavioral and Mental Health researchers empirically investigate mechanisms of SUDs, and develop and evaluate novel interventions to reduce harm and to support positive health outcomes. Behavioral and Mental Health practitioners, clinicians, and social workers implement interventions and adapt them to the specific populations to affect health outcomes. These interventions or the process of their implementation may include computational support in forms of digital artifacts, information management approaches, and more.

Each of these lenses connects to a number of related areas that influence the nation's response to SUDs and none of them work in isolation. For example, Ubicomp sensors may be leveraged to generate data analyzed by the informatics researchers. Insight from this analysis may inform future behavioral and mental health interventions, as well as public health and policy directions. Unfortunately, despite the significant potential for synergy and collaboration between the people who employ these lenses in their work, there are a number of barriers that may prevent meaningful collaboration, such as: (1) institutional siloing of research areas, (2) disciplinary differences in values and what constitutes an intellectual contribution, and even (3) the scope and source of funding traditionally sought for particular type of work.

The Computing Community Consortium (CCC) sponsored a two-day workshop titled *"Computational Support for Substance Use Disorder Prevention, Detection, Treatment, and Recovery"* on November 14-15, 2019 in Washington, DC. The workshop's goal was to bring together an interdisciplinary group of leading researchers and practitioners to identify opportunities and challenges for enabling innovative computational support for prevention, detection, treatment, and long-term recovery from SUDs.

As outcomes from this visioning process, we identified three broad opportunity areas for computational support in the SUD context:

1. Detecting and mitigating risk of SUD relapse,
2. Establishing and empowering social support networks, and
3. Collecting and sharing data meaningfully across ecologies of formal and informal care.

We expand on each of these areas in section 3, identifying prior work in the space, challenges and research gaps in the area, and specific opportunities for computational support.

Throughout this visioning process, the group also identified cross-cutting challenges that affect how SUD computational support research is planned, carried out, and disseminated:

1. Ethical considerations for working with stigmatized and vulnerable populations,
2. Identifying and managing privacy risks and concerns data collection,
3. Identifying and reducing potential harm when deploying computational SUD interventions,
4. Enhancing theoretical underpinnings of interventions and how they may fit into broader theories of change,
5. Recognizing and responding to the disproportionate burden of SUD in specific disparity populations, and
6. Implementing, transferring, and sustaining interventions to create a path for innovative technical work to influence practice.

We discuss each of these six challenges in section 4. Based on the collective expertise and experience of workshop participants, we articulate each challenge and propose initial directions for addressing it.

## 2. Interdisciplinary Visioning Process

This workshop brought together scholarly leaders (both researchers and practitioners) who represented one or more of each disciplinary lens (see Section 7 for complete participant list). Given this diversity of backgrounds, we built a shared understanding by articulating the priorities of our area through lightning talks and by holding two panels sharing both lived experience and clinical perspectives on SUDs. Following this process, participants were assigned to one of seven interdisciplinary teams and followed the IDEO ("Innovation Design Engineering Organization") ideation process [25] to each articulate more than 50 research challenges, problems, and open questions where computing has an opportunity to address SUDs. Each team included at least one person who was currently in active recovery from SUDs to help vet and guide ideas based on their lived experience. We informally discussed ideas as a larger group and reconvened the next day to review them by culling, combining, and clustering the most promising research ideas. Finally, each group picked one such opportunity or challenge to expand and present back to the larger group.

Based on these presentations and ensuing discussions, we identified three overarching opportunity areas. We articulated each opportunity area by describing relevant prior/current work, identifying specific computational directions, and highlighting cross-cutting challenges.

## 3. Identified Research Opportunity Areas

Through our visioning process (see section 2), we identified three specific promising areas for interdisciplinary work where computing can provide a unique perspective, approach, or set of methods.

## 3.1 Detecting and Mitigating Risk of SUD Relapse

Relapse is a common part of the SUD recovery process — with 75% reporting a relapse in the first year of recovery (rates similar to other chronic health conditions) (e.g., [35,45]). There is an opportunity for computing to provide a better empirical understanding of the behavioral causes of relapse, to mitigate its impact through early detection, or to prevent relapse altogether by helping individuals minimize negative and maximize positive recovery behaviors.

### 3.1.1 Sensing Risky and Healthy Behaviors

Outside of computing, most prior work quantifying risk of relapse focuses on patient history (e.g., comorbid disorders, prior trauma, impulsivity) (e.g., [40]) or retrospective self-report of behavior (e.g., [45]). Computational support provides an opportunity to gather more reliable data about risky and healthy behaviors and associate this data with a more valid ground truth on relapse. For example, mHealth approaches allow gathering self-reported behavioral or cognitive states more reliably through ecological momentary assessment (EMA) [37], the integration of biophysical sensor data such as heart rate and galvanic skin response [4], and inclusion of environmental variables such as location through geofencing [16]. Work combining multiple such approaches shows great promise towards understanding the contextual determinants of drug use (e.g., [3]). Unfortunately, substantially less research has focused on detecting positive or healthy behaviors that may mitigate risk of relapse, such as engaging with mindfulness activities or other recommended forms of treatment. Mobile sensing also allows researchers to gather more reliable measures of ground truth regarding relapse without needing to rely on self-report. For example, alcohol consumption may be detected using existing sensors embedded in mobile phones [1], or devices may be augmented with additional biochemical sensors to periodically check for use of specific substances (e.g., [62]).

However, most of the current work in this area is fairly preliminary, with four major intellectual challenges that need to be addressed before it can have a significant impact on SUDs. First, many mHealth approaches currently place a high burden on the user (e.g., answering EMA prompts, providing biochemical samples, charging devices), so they **rely on effective incentive strategies and patient acceptance** (see 3.1.2). Second, most passive biophysical data collection approaches currently **lack the robustness necessary for clinical deployment,** let alone transfer to practice. Third, effectively combining data from multiple diverse sources relies on sophisticated and robust algorithms that can provide **accurate personalized prospective quantifications of risk even in the face of heterogeneous, noisy, or missing data.** Finally, even if risky or healthy behaviors can be detected accurately and in ways that are acceptable and feasible, it is an open challenge as to **how this information may be presented in actionable ways to relevant stakeholders** (see 3.3.2). All four of these challenges require work across interdisciplinary teams that include significant computational expertise.

### 3.1.2 Incentive Structures Across Timelines of Recovery

Motivating engagement with SUD data collection, acceptance, and compliance with recovery treatment is a major challenge in both detecting and preventing relapse. Early positive extrinsic incentives (e.g., micropayments [43], contingency management [50] of various forms) may be effective and have been used in clinical studies and transferred successfully to industry (e.g., WeConnect Recovery App[1]). While only somewhat effective, coercive punitive measures (i.e., "prison or treatment") for treatment compliance may also be effective in the short term if delivered through a well-structured comprehensive program [13]. However, it is both impractical and ineffective to continue providing extrinsic contingency management incentives indefinitely [50]. Individuals must develop personal intrinsic motivations to stay in recovery, which may be negatively impacted by the use of initial extrinsic incentive structures [59]. Promising approaches to helping participants develop intrinsic motivation include therapeutic techniques like "acceptance and commitment therapy" and "motivational interviewing," which may be

---

[1] https://www.weconnectrecovery.com/

**3**

delivered as part of traditional treatment [50] or may be supported by telehealth approaches and delivered remotely [14]. Additionally, helping people integrate and reflect on positive or pleasurable activities in recovery can aid relapse prevention and provide intrinsic motivation to continue recovery [29]. Incentives for engagement may also consist of individual benefits such as an opportunity to engage with something beautiful or pleasant to the senses (e.g., identified as an engagement strategy in participatory design with youths [61]), pointing to opportunities for immersive multi-sensory technologies (e.g., VR, digital arts, haptics) as components of interventions. Finally, long term intrinsic motivation may be increased by identifying opportunities for social incentives, such as mentoring newcomers to recovery [63] or forming social support networks with others in recovery (see 3.2.1).

There are a number of intellectual challenges in providing incentive structures across timelines of recovery. The first challenge in effectively providing incentives is that they **rely on participants choosing to share accurate data.** This may be problematic in cases in which incentive structures may be external (e.g., condition of probation) and foregrounds substantial ethical considerations regarding personal agency, privacy, and control over increasingly intimate (e.g., heart rate, biochemical excretions, etc.) data. The second challenge is knowing **which incentive structure to provide at which time along the continuum of recovery** or how to combine incentive structures to achieve the best effect for a particular individual (given that it is difficult to effectively measure ground truth on motivation as an outcome variable), particularly when transitioning to a longer term intrinsic motivation model. The third challenge is **addressing the right level of behavioral granularity.** Traditional measures of abstinence from substance use (or even reduction in use) may be too coarse to be actionable for individuals in recovery. For example, it would be better to signal to a patient that some behavior is risky and might lead to using in the near future, rather than to tell them after a relapse has occurred. Thus, a major challenge to the implementation of effective incentive structures is understanding which behaviors to incentivize and disincentivize, and how (see section 3.1.3).

### 3.1.3 Helping Individuals Avoid Risky and Approach Healthy Behaviors

Assuming that we can reliably collect and interpret data about risky and healthy behaviors, there still remains the open question of how such data may be used to influence individual behavior. An individual in recovery makes a myriad of choices every day, some of which may increase and others decrease risk of relapse. Computational support may help individuals become aware of the risks associated with particular behaviors and may provide persuasive technologies (e.g., [58]) and just-in-time adaptive interventions (JITAIs) (e.g., [28]) to nudge them towards healthier practices. Such systems may include mechanisms for increasing relevance of information offered to individuals (e.g., through tailoring, personalization, feedback), attempt to enhance motivation to engage in positive behaviors (e.g., through rewards, recognition, competition), or trigger an escalation to a "high touch" intervention (e.g., social facilitation, clinical support) [58]. Finally, there are a number of cases where healthy behaviors may need to be taught rather than simply advocated. For example, there are mindfulness techniques that may provide opioid-free ways of managing chronic pain [24]. Novel computing technologies such as telehealth and virtual reality guided courses may scalably disseminate these types of training without sacrificing fidelity. In fact, it's possible that the immersive, multisensory experiences made possible by these technologies may increase engagement (especially when these environments are designed with careful participation from the arts) and improve instruction/training.

One major challenge in this endeavor is the substantial **gap between the growing technological capabilities for delivering a variety of just-in-time or persuasive interventions and the research on how these interventions should be developed and evaluated** [39] in clinical and community settings. Many persuasive technologies fail to clearly specify the model and intended mechanisms of change [58]. Existing theoretical models of change may not adequately account for these systems, their effects (particularly possible unintended effects), or how their capabilities may be leveraged to enhance SUD recovery. The development of new models will require work across interdisciplinary teams to ensure that these theories are both evidence-based and sufficiently generalizable to account for emergent forms of technology. Another major challenge is that effective responses to risky and healthy behaviors rely on **holistic understanding of an individual and accurate interpretation of their context.** In cases of tailored interventions, an understanding of context may be the difference between intervening "just-in-time" (e.g., as an individual tries enters a high risk geo-fenced area with an intention to use) versus triggering a risky behavior (e.g., suggesting to an individual that this is an area where others frequently go to use). In case of providing specific training to support healthy behaviors, the educator or system must be able to understand how a person is engaging with new positive behaviors in the moment and in their daily life in order to support their effective use. Current technologies provide narrow snapshots of a participant's perspective through specific sensors and streams of data, but no current systems support this level of holistic understanding as it requires rich, multimodal data interpreted by personalized expert-in-the-loop algorithms.

## 3.2 Establishing and Empowering Social Support Networks

Addiction has been described as the "disease of isolation," emphasizing the need for social connection as an important component of recovery [20]. Indeed, people who receive appropriate social support are more likely to get into treatment (e.g., [43]) and have better treatment outcomes (e.g., [12]). There are a number of opportunities for computing to empower and enhance social support networks for individuals with SUDs.

### 3.2.1 Helping Individuals Find "Their People" or "Family of Choice"

Social support, whether from family, friends, 12-step programs, or other sources is a critical predictor for long-term recovery (e.g., [4,12]). However, many people enter recovery without such existing social networks in place or with ones that are inherently entangled with a traumatic past [7]. Twelve-step groups and twelve-step facilitation are cost-effective approaches for providing effective peer support to improve SUD outcomes [26], and for many begin to constitute a "family of choice" in recovery. However, it may be difficult for individuals to find the groups that are "right" for them, particularly in cases

---

**Case Example of Technology Use for SUD Recovery during COVID-19 Pandemic.** While occurring after the workshop event, at the time of writing this report, the COVID-19 pandemic is raging and its effects are being felt in all areas of society, including among people with SUDs. Given the timing of the workshop and of this report, we mention the relation of the two as a brief case example, but we expect the connections will be complex and profound.

As noted in Section 3.2, addiction has been described as the "disease of isolation," making the primary measure to control the spread of COVID-19—social distancing—particularly problematic for people with SUDs (e.g., [2,21]). Already in the course of the pandemic there has been a decrease in the number of people seeking treatment [2], while at the same time existing online recovery meetings have seen an explosive growth, and many in-person meetings have temporarily moved to an online format or struggle provide services in-person (e.g., [55]). We expect that there will be important research investigations addressing SUDs and the COVID-19 pandemic, including its effects on people in active addiction, people in early or long-term recovery, caregivers, family, and the community. There is also the opportunity to examine the large-scale transition of people from in-person meetings to online recovery meetings, and factors that supported or impeded this transition. Long-term, there may be a shift in how people choose to seek support as infrastructure for online meetings becomes more readily available and there may be a greater recognition of the barriers to online participation faced by people for whom physical isolation may be a daily part of life (e.g., people with disabilities, immuno-compromising health conditions) [48]. We anticipate that the COVID-19 pandemic will permanently change the landscape of recovery services and support. The kinds of online tools and communities discussed in this report will likely serve as a critical part of that change.

in which they may represent marginalized demographics such as indigenous or LGBTQ people (e.g., [18]) or live in disadvantaged areas [51]. Beyond finding a group, finding the right dyadic mentorship, sponsorship, or coaching relationship can be beneficial but a substantial challenge for somebody in early recovery [21]. This foregrounds opportunities for computational support to increase the diversity of networks available (e.g., [40]), facilitate access to these networks [39], and match individuals with group (e.g., [47]) or dyadic support (e.g., [21]) that may best meet their needs.

It is also important to acknowledge that many individuals who use substances may not identify with some models of addiction, abstinence-focused programs, or other commonly-available treatment approaches. These individuals may find "their people" in online communities in which they seek treatments and harm-reduction supports (e.g., [8,10]). Computational approaches may provide a scalable way for the medical community to engage with, monitor outcomes, and remain aware of these treatments and their effects [8]. These spaces could also allow providers opportunities to generate new knowledge on treatment and then promote evidence-based strategies in a manner that is more community-engaged and respectful of individuals seeking treatment and support. These online forums may then provide an opportunity to engage in remote participatory design (e.g., [30]) or community based participatory research with groups who may currently not be reached or feel disenfranchised by conventional treatment approaches.

A major challenge in this area of research lies in **evaluating social support interventions and systems, particularly with an eye for potential harm or unintended consequences.** For example, most of the work with alternative treatment communities is preliminary and it is not clear in which cases the support or information these communities provide leads to positive outcomes. There are many examples of online communities that encourage harmful behavior and for which thoughtful moderation would be more appropriate than encouragement (e.g., [9]). Additionally, even effective social support programs may lose efficacy when transitioned into computationally-mediated environments — loss of connection [48] and anonymity [41] are two examples of concerns in this space. Meaningful evaluations in this space must not be an afterthought. Instead, researchers should consider evaluation throughout the design process, integrate community feedback meaningfully into both the system and the evaluation study design, and iteratively consider the effects of the systems on its users and non-users during and after a deployment.

### 3.2.2 Opportunities for Mutual Support by Caregivers

While social support is critical to successful recovery, those who provide that support may face significant challenges in doing so. When the caregiver is a family member, they may face substantial distress and burden [44]. While a number of programs (e.g., Al-Anon) and interventions (e.g., CRAFT) exist to provide support for family members of people struggling with SUDs, caregivers need help in identifying and accessing the right support program [17]. Computational strategies we describe in 3.2.1 for helping individuals with SUDs may also be applied for helping caregivers find "their people." Outside of the family context, peer support providers, such as sponsors in 12-step programs, provide a compelling and accessible network of support for people in recovery. However, this type of mentorship can be a substantial responsibility, and sponsors rely on adapting a variety of sociotechnical practices to help [21]. Caregivers may experience "burnout," isolation, and struggle with boundary-setting. This points to opportunities for the design of sociotechnical approaches that enhance mutual social support and information exchange among caregivers. Online communities may provide a unique opportunity for this kind of support because they bypass issues of geographic proximity and may allow pseudo-anonymous support seeking. Such approaches have been successful in the past for other caregivers of chronic condition patients (e.g., [13]). One challenge in both caregiver and patient contexts is in integrating evidence-based support into technical systems which have to be designed or adapted for these approaches (e.g., [24]). This requires teams that have both the technical expertise to be able to build novel systems and the clinical expertise to be able to guide system design and facilitate the support interventions once they are in place.

### 3.3 Collecting and Sharing Data Meaningfully Across Ecologies of Formal and Informal Care

The Bronfenbrenner Ecological Model (see Figure 1) provides a useful lens for patient-centric ecological approaches to SUDs [24]. A person struggling with an SUD is at the center of an ecosystem of people and institutions — cultural and societal factors on the outside, institutions like communities and social services closer, and immediate support like family, friends, doctors, and teachers constituting the innermost circle. Computational support provides opportunities to bridge across these elements and systems to help people struggling with SUDs.

**Bronfenbrenner Model Applied to SUDs**

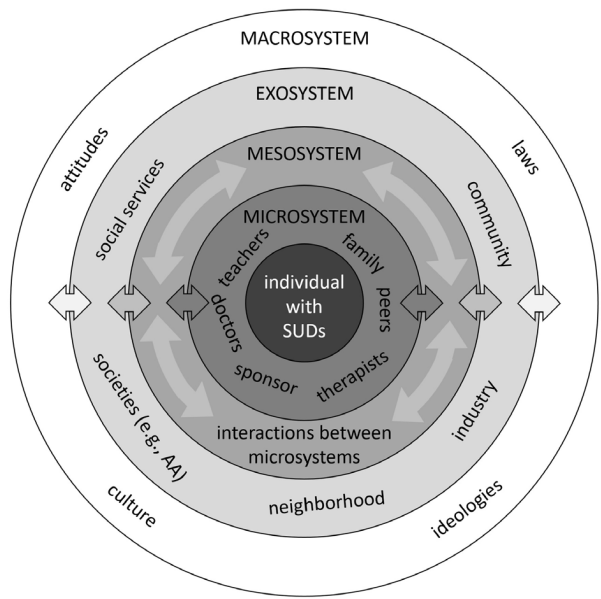

*Figure 1: Four concentric circles describing the macro-, exo-, meso-, and microsystems that may affect an individual struggling with SUDs.*

### 3.3.1 Sharing Data Across a Patient-Centric Network to Support Individuals in Crisis

Perhaps the most critical time to mobilize support across the entire ecology of care is when an individual is facing an immediate or imminent crisis (overdose, relapse, or high risk of developing an SUD). Computational support has been effectively leveraged with specific elements of the ecological microsystem (e.g., peer support [40], family [50], clinicians [23,27], counselors [10]) and exosystem (e.g., mobilizing emergency community response to overdoses [1]) levels, but may be lacking at the mesosystem level of enabling connection between these individual elements. For example, there are no existing mechanisms to support family members who want to communicate with a 12-step sponsor, anvd either one may face challenges if trying to connect with a treatment provider or doctor. Some of these boundaries represent meaningful constraints to protect confidentiality or anonymity, but others become problematic during moments of crisis where immediate action may save lives and each element of the patient-centric social network may hold important contextual information.

A critical challenge in this area is being able to **reconcile and design for the potentially divergent values of the many stakeholders** (e.g., individuals in crisis, family members, peer support group members, health professionals, community members) who may be involved in a potential patient-centric network. For example, one critical value may be "privacy" or "agency," as one person's "support" may feel like "surveillance" to another person. Integrating such systems into clinical practices presents substantial additional challenges around acceptance, fidelity, and patient confidentiality [34]. This points to the importance of developing such systems using participatory approaches with multiple stakeholders involved, where methods and paradigms such as Value Sensitive Design may be particularly helpful [5] or new methods may need to be developed to design for ethical data sharing within these complex ecologies.

### 3.3.2 Making Data from Multiple Sources Actionable for Care Teams

Computational support can facilitate data collection from a myriad of data sources about any particular individual. From social media use, we can determine factors like risk of depression [9] or severity of particular mental health symptoms (e.g., [5]), including substance use [57]. From on-body sensors, we can gather biophysical measures, such as stress (e.g., [23]) or immediate substance use (e.g., [18]). Consumer mobile devices and internet-of-things audio/video sensors in the environment can also provide increasingly detailed passive sensing about location, activity, and substance use (e.g., [1,44]). The wide availability of consumer mobile devices also provides easier in-the-moment access to self-reported measures for tracking aspects like stress, mood, cravings, and specific behaviors (e.g., [27]). Beyond the individual, various forms of electronic records may empower others in a support network, health system, or community to provide information about the behavior, state, or clinical history of an individual with SUDs. Furthermore, while most work has focused on using single sources of data, combining multiple modalities frequently leads to emerging insights unavailable from separately-considered sources (e.g., [3]). Computational support may generate a wealth of data potentially available to care teams, but meaningfully using this data presents two substantial challenges.

The first major challenge and constraint in this research opportunity is in **combining data from this large variety of sources.** There are obvious technical challenges to combining multiple data streams with different granularity, data quality, and completeness, though we are making strides to address some of these (e.g., [23]). Certain technical challenges associated with combining disparate data can be mitigated by adopting data standards and terminologies. For instance Health Level 7 International[2] is an accredited not-for-profit standards development organization that focuses on electronic health record data to support patient care. More fundamentally, there are also privacy, norm, and access constraints to whether and how some forms of data may be collected (e.g., AA meeting attendance) and shared (e.g., EHRs). To address some of these constraints, the patient may serve as the collector and curator of self-generated data and electronic records [12]. In other areas of health, such data has been shown to serve as a boundary object between patients and providers, facilitating collaboration and decision-making [5]. However, the context for SUD recovery may be sufficiently different that it needs separate consideration. Researchers should consider the values of each stakeholder group and the various power dynamics—perceived or actual—that inform the practices and culture within and between stakeholder groups.

The second major challenge is the need to develop computational systems to **support data/dimensionality reduction, rationalization/explanation, and visualization to mediate between the massive diverse raw streams of data available regarding a given patient and the limited capacity, time, and importantly, the demands and expectations of the care team.** This presents substantial challenges both from the perspective of developing algorithms which are capable of this reduction (without introducing substantial bias into the process) and in terms of the information visualization approaches used to present the data to various stakeholders. The level visualization or explanation necessary may be different depending on the type of collaboration or decision-making it is meant to support. Developing such systems will require technologists, people in recovery, and various members of the care team to work closely together in long-term partnerships, rather than a sequential model of system development.

### 3.3.3 Community-Level Sensing and Insights

Beyond sensing data about any particular individual, computational support may provide an opportunity to understand aggregate patterns of substance use and recovery from social media traces or volunteered data, such as population-level mental health trends [8], emerging substances or practices [6], and resources available in particular geographic locations (e.g., BMLT Narcotics Anonymous resourcing[3]). Collecting and aggregating such data supports policy and community-level decision making, for example by identifying underserved communities or endeavoring to provide more relevant and timely resources to groups, particularly those who might be under-represented in conventional epidemiological studies. Such aggregation could also promote getting information about relevant issues and resources (e.g., 12-step meetings, needle exchanges, community services, homeless shelters, domestic violence support) to the individuals who would most benefit from them through tailored person-resource matching.

One technical challenge in this area is in **gathering and aggregating data from multiple sources,** many of which may be incomplete or outdated and some of which may be hyper-local in nature and content. While there have been some initial successes in this space by combining information retrieval and crowd work approaches [47], much work remains before such systems are robust and complete. A second, more fundamental, critical computational challenge in this area is **respecting individual privacy while gathering and sharing sufficient data to support meaningful aggregate analysis.** Many individuals may be unaware how broadly-aggregated data about them is used by companies or researchers (and in fact, may object to such practices if made aware of them [15]). Research projects in this space must tackle the thorny ethical challenge of balancing individual desires and societal benefit, as well as understanding whether and how consent may be given in the context of community-level data collection. If such work is worth pursuing, it would need to take special care to protect individuals in aggregated data sets from the potential consequences of deanonymization (which is particularly critical in this context, where such deanonymization could have substantial legal, financial, personal, or professional consequences for individuals). The computational approach of "differential privacy" may provide a compelling way for groups and organizations to protect participants while sharing aggregated data [11], though it may need to be adapted given the unique considerations of SUDs.

---

[2] www.hl7.org

[3] https://tally.bmlt.app/

## 4. Cross-Cutting Challenges and Considerations

Beyond the specific areas of research opportunity, workshop participants identified a number of cross-cutting challenges and considerations for computational support in the SUD context. In this section, we articulate these considerations and outline initial steps toward solutions.

| Challenge | Opportunity |
| --- | --- |
| Ethical considerations for working with stigmatized and vulnerable populations in the SUD context. | Convene a group to publish a set of guidelines and best practices to be disseminated to institutional review boards.<br><br>Embrace a participatory research approach that includes critical stakeholders as advisors throughout each research phase, including budgeting for their time and investment. |
| Identifying and managing privacy risks and concerns regarding SUD data collection. | Develop an empirical understanding of, and build infrastructure for, (1) people to evaluate their privacy risks and benefits in both commercial systems and research investigations, (2) technical approaches that provide stronger assurances of individual privacy, and (3) policies and systems for simplifying post-hoc removal of data. |
| Identifying and reducing potential harm when deploying computational SUD interventions. | Establish peer and community review processes that allow researchers to obtain external vetting of deployment strategies before proceeding (i.e., similar to the way that clinical researchers may publish their protocols) and review the project at multiple points during and following the deployment of an intervention.<br><br>Emphasize and support proven evidence-based practices when possible, and encourage ongoing research when evidence is unavailable or unclear. |
| Recognizing and responding to the disproportionate burden of SUD in specific disparity populations. | Use sensing technologies to understand mechanisms driving disparities in SUD prevalence and outcomes in disparity groups.<br><br>Develop computational support and intervention models that are targeted to populations that experience disparities in SUD prevalence or outcomes.<br><br>Evaluate impacts and potential differential interactions with one-size-fits-all SUD interventions (i.e., intended for population-wide implementation). |
| Enhancing theoretical underpinnings of computational SUD interventions and how they may fit into broader theories of change. | Inform computational interventions for SUDs with accepted theories of behavior change by including experts in these theoretical perspectives on project teams from the outset of the work. |
| Implementing, transferring, and sustaining interventions to create a path for innovative technical work to influence practice. | Create and expand funding calls and networking opportunities that support long-term interdisciplinary work and transitions at critical stages. |

COMPUTATIONAL SUPPORT FOR SUBSTANCE USE DISORDER PREVENTION, DETECTION, TREATMENT, AND RECOVERY## 4.1 Ethical Considerations for Working with Stigmatized and Vulnerable Populations in the SUD Context

People living with SUDs often face substantial stigma in society. Many of them also face the intersectional challenge of being a member of a vulnerable population or marginalized group. Researchers must employ special considerations and potentially adjust practices to reduce potential legal, financial, professional, and personal harms to participants. At the operational level, this may include adjusting practices for documenting consent (e.g., IRB waiver of documentation of consent), adjusting human subjects compensation to support anonymity, and providing stronger confidentiality guarantees (e.g., NIH certificate of confidentiality). Current practices for participant protection seem to vary by institution and funding body, so an initial first step may be convening an interdisciplinary working group to document best practices to share with researchers and institutions.

At the broader methodological level, it is also clear that researchers may not always have a reliable intuition for what constitutes harm and how their work may relate to the lived experiences of people struggling with SUDs. One opportunity to address this may be to embrace participatory research approaches (e.g., participatory design, community-based participatory research, cooperative inquiry) as ways of **engaging with members of recovery communities throughout the research process.** Two critical considerations for creating such partnerships may be inclusion and reciprocity. Inclusion can only be achieved by continually considering people and groups that may not have a "seat at the table" (members of economically disadvantaged groups, underrepresented minorities, people with SUDs who do not identify with traditional recovery communities, etc.) and active outreach and commitment of resources to include such groups. Reciprocity is necessary to support long term engagement with communities and groups by equalizing power, confronting science's colonial impulses, and prioritizing relationship-building rather than transactional interaction in community-engaged research. At the very least, this must include concerted efforts to give back to the partner communities in ways that *they* value, as well as disseminating research findings back to these groups in appropriate ways. A broader challenge is that all of this engagement is only possible if agencies and institutions recognize and prioritize the relational aspects of doing SUD work well.

## 4.2 Identifying and Managing Privacy Risks and Concerns Regarding SUD Data Collection

A specific ethical consideration for working with people with SUDs is in identifying and managing privacy risks and concerns of data collection. Most promising computational directions focus not just on collecting specific information, but rather on also collecting streams of data across multiple modalities and granularities and then inferring new variables of interest from those sources. This makes it more difficult for participants to consider risk and provide truly informed consent, as it is not always clear that seemingly innocuous forms of data (e.g., phone accelerometry, GPS signal, linguistic content of online posts, audio gathered in the environment) can be used to infer potentially sensitive information such as substance use, risk of relapse, or membership in anonymous communities. Furthermore, it is not always clear to the users of technologies which data is gathered and who has access to it or the insights inferred from it (and thus the consequences of sharing such data). This can be particularly pertinent in the case of for-profit online tools (e.g., Facebook), many of which serve host to recovery communities but may not include any protections for individual anonymity against data being sold to health insurance companies or against surveillance by law enforcement.

We identified three major opportunities in this space, all requiring substantial research and infrastructure to be viable. First, there is a need for **understanding and providing clear and timely explanations to study participants and system users regarding data** to be collected, dissemination of such data, and insights to be inferred from it. Second, there is a need for **stronger assurance of individual privacy protections in aggregated data.** The area of differential privacy provides a promising approach but may benefit from more focused collaboration with human-centered technologists and with domain experts in this sensitive context. Finally, there must be **ways for participants or users of systems to** address data privacy issues post-hoc, which can include requests to be removed from any particular dataset. While many systems already provide some measure of this to comply with European Union GDPR (General Data Protection Regulation) practices, in this sensitive context such protections must be extended more broadly and must be demonstrated to be clearly legible and accessible to the target audience.

## 4.3 Identifying and Reducing Potential Harm when Deploying Computational SUD Interventions

Incentive structures for researchers and industry innovators frequently prioritize demonstrating measured benefits in a self-contained study over more holistic and long-term evaluations of computational interventions. However, there are many potential ways that an intervention may introduce unanticipated effects or substantial harms either through the nature of the intervention itself or through the ways that it is delivered. First, computational interventions may have unintended negative effects on existing practices. For example, people in recovery communities commonly worry about computational interventions excluding people without access to technology, replacing in-person connection with more shallow online tools, violating anonymity, and interfering with building individual autonomy and agency [60]. Researchers need to understand such concerns (which may be a substantial task, given that some terms may not be operationalized and considered in the same way by researchers and people in recovery [49]) and ensure that these are included in the empirical evaluations of interventions. Given the high-stakes nature of the SUD space, it is also important that these harms are evaluated throughout the deployment and that there is a clear contingent strategy to provide additional support (e.g., off-boarding, referrals to other services, professional counseling support) for participants in cases where substantial harms may be identified during the study. Second, computational interventions increasingly rely on non-transparent algorithms (e.g., to classify behavior from sensor signals, to create estimates of relapse risk, to socially match people for support). Such algorithms may be prone to reproduce existing bias in the training set or succumb to feedback loops where minor differences in the data set are amplified by the algorithm over time [41,56]. The presence and effect of such potential algorithmic harms may not be clear at the point of initial deployment. Thus, researchers must provide guidance for periodic auditing of algorithms for bias, output interpretation, and uncertainty estimation for those using algorithmic results in either system design or decision-making. Third, in cases where a computational intervention is shown to provide substantial benefit, participants may be harmed by the intervention being removed or becoming unavailable at the end of the study. As such, it is critical for researchers to consider alternative ways of providing similar benefits and provision for other forms of safety net while off-boarding participants.

There are several potential approaches to identifying and mitigating the potential for such harms. One is a growing movement in computing to engage peer review as a process for encouraging researchers to consider and evaluate potential harms of the systems they build (e.g., [19]). However, doing so at the publication stage may be too late. The research community must consider mechanisms for introducing peer review at earlier stages in the process of designing computational interventions and their deployments. One approach may be introducing a new publication category inspired by "trial protocol papers" in the medical domain [30], allowing a design and deployment strategy to be peer reviewed and made public prior to carrying out the empirical study. Additionally, given that peer reviewers may not have the necessary insight to understand some potential harms from the point of people with lived experience with SUDs, a stronger review approach would also include some form of "community review" in addition to traditional scholarly peer review.

## 4.4 Recognizing and Responding to the Disproportionate Burden of SUD in Specific Disparity Populations

SUDs disproportionately affect specific populations, such as members of rural communities, Indigenous people, and LGBTQ people, yet they may be poorly served by prevailing models of treatment and recovery for SUDs. For example, LGBTQ people have significantly elevated odds of lifetime substance use than heterosexuals and those whose gender identity matches that which they were assigned at birth [7,32]. These disparities have been linked to high rates of bullying, violence, stigma and discrimination experienced by these groups [33]. LGBTQ people may also experience biases, discrimination and lack of cultural competence when seeking mainstream substance use disorder treatment [31,54]. Furthermore, LGBQ people report lower levels of satisfaction with substance use treatment than heterosexuals [52]. Due to these challenges, the Institute of Medicine prioritized development and evaluation of substance use interventions for this group [22]. More recently, a small number of digital interventions have demonstrated the feasibility and promise of digital substance use interventions with LGBTQ youth (e.g., [51]) and adults (e.g., [46]). Addressing the unique needs of different disparate populations poses specific challenges in understanding the mechanisms driving these disparities and developing and evaluating interventions with these groups.

We identified three major opportunities for addressing these challenges. First, there is potential for **using sensing technologies to understand mechanisms driving disparities in SUD prevalence and outcomes in disparity groups.** For example, biopsychosocial factors such as minority stress in LGBTQ populations might be helpfully assessed using sensors that measure physical and emotional states. There is also an opportunity to **develop tailored computational support and intervention models that are targeted to populations that experience disparities** in SUD prevalence or outcomes. It is also important to partner with others to **evaluate impacts and potential differential interactions** with one-size-fits-all SUD interventions (i.e., intended for population-wide implementation). For example, people with co-occurring mental health disorders and/or trauma histories may respond differently to SUD interventions than other groups. It is critical for computing researchers to understand and address disparities to ensure that SUD interventions do not widen SUD-related disparities.

### 4.5 Enhancing Theoretical Underpinnings of Computational SUD Interventions and Their Fit into Broader Theories of Change

Computing fields and funding mechanisms often prioritize technical novelty and innovation at the expense of other elements of a computational intervention. For example, systematic reviews of the area of "persuasive computing" (which is often applied in health contexts to influence behavior change) found that most papers did not specify any kind of model or route of behavior change [58]. This contrasts heavily with more clinical perspectives and NIH approaches to research, where the model of change is typically a critical component of both intervention design and evaluation planning. In health domains, a number of theories of behavior change combine decades of empirical insight and multiple more specific models into overarching frameworks to characterize and guide interventions, with two most commonly-used ones being the transtheoretical model of health behavioral change [42] and the COM-B behaviour change wheel model [36]. However, while both of these models include dozens of constructs and critical assumptions, neither of them explicitly addresses the potential role of tools like technologies. On the other hand, while HCI does employ broad theories of how technologies may be leveraged in social contexts (e.g., distributed cognition, activity theory [17]), they are infrequently used and not specific to the health domain.

We identified two major opportunities to address these theoretical gaps. First, technologists must recognize that **for computational support to be relevant and meaningful for translation to clinical interventions, it must be informed by theories that are recognized within health domains** from project outset. This may be best achieved by doing computational intervention design within interdisciplinary teams that include researchers well-versed or expert in such theories. The second opportunity is leveraging the peer review process to encourage computational research which advances and expands current theories of change to include the role of technology and provide clear operational guidance to technology designers. This may include the **development of new theories of computational interventions or carrying out empirical work that demonstrates how technology fits within the core constructs or assumptions of current models of change.**

### 4.6 Implementing, Transfering, and Sustaining Interventions to Create a Path for Innovative Technical Work to Influence Practice

Throughout the workshop, it became clear that while all of us shared a common goal of improving the practices of SUD prevention, detection, treatment, and recovery,

**12**

our discipline and domain determined how we sought to have this impact. For example, contextualizing within the NIH Stage Model for Behavioral Intervention Development [64], almost all work from computing domains focused in the first two stages — informing intervention design and generating new interventions. There was a clear "seam" in moving from creating novel computational interventions to establishing their efficacy in controlled trials. Similarly, clinical researchers expressed frustration with a similar "seam" between establishing pure efficacy in trials versus real-world efficacy and effectiveness in real communities. While all of us sought to do work that moved towards eventual implementation and dissemination in community settings, we each faced disciplinary and funding barriers in that endeavor. It is apparent that for us to benefit from each others' expertise we must have **mechanisms to support long-term team science.** Such support mechanisms are particularly critical at the "seams" of stage transitions, including moving from empirical work or theory to a computational intervention, from technical proof-of-concept to robust prototype, from prototype to piloted intervention, from pilot to clinical trial, and from clinical trial to implementation in practice. Support for building such interdisciplinary bridges (see 5.2) must be both relational (e.g., workshops, events, and consortia) and operational (e.g., funding mechanisms).

## 5. The Future of Computational Support for SUDs

Based on the identified research areas and cross-cutting themes and considerations, we describe disciplinary and interdisciplinary opportunities for the future of computational support addressing SUDs.

### 5.1 Opportunities Within Disciplinary Domains

Computational support for addressing SUDs benefits from research in a number of disciplinary areas, as well as providing new opportunities for fundamental scientific research in those fields. Returning to the four disciplinary lenses identified in the introduction:

▶ **For Human-Computer Interaction (HCI) research,** the SUD context provides a unique perspective on a number of core research areas within HCI. These include (1) designing and evaluating computer-mediated communication and social recommender systems to enable establishing and empowering support networks for both general and disparity populations, (2) informing and developing persuasive and personalized technologies for training, supporting, and incentivizing behavior change that are sensitive to holistic individual, social, and cultural contexts, and (3) developing and validating new methods for community-driven and participatory design to acknowledge and reconcile divergent values between multiple stakeholders in the SUD context.

▶ **For Ubiquitous Computing (Ubicomp) research,** computational support for SUDs provides a critical context which benefits from combining behavioral, biophysical, self-report, and social media data in novel ways. Fundamental scientific research challenges include (1) the development of novel sensors and algorithms for inferring high-level contextual constructs (e.g., "risk of relapse") that are relevant to SUD disparities, interventions and their evaluations, and (2) investigating human-centered approaches for making such data understandable and actionable for individuals, caregivers, and teams.

▶ **For Informatics & Data Science research,** the rich heterogeneous sources of data collected about the states and behaviors of individuals with SUDs offer a compelling context to address fundamental data science challenges, including (1) adopting data standards and terminologies to facilitate sharing data and maintain semantic properties, (2) developing algorithms to meaningfully combine, reduce, visualize, and explain such data to relevant stakeholders, and (3) improving data ethics through advances in usable privacy and human-centered approaches to differential privacy.

▶ **For Behavioral and Mental Health research,** computational support offers a new set of tools to create and expand access to successful treatments and interventions and to evaluate interventions in a more granular way both at the individual and at community levels. The major fundamental research opportunities are in (1) carrying out empirical and theory-building work to integrate and address technology's role in models of change, (2) guiding the design and deployment of novel computational systems and interventions, and (3) informing, auditing, and interpreting the output of algorithms to extract insights and inform decision-making across ecologies of care.

However, the most important opportunity identified through the workshop was not the potential for any given discipline to contribute but rather the importance of interdisciplinary collaboration and coming together to address the broader cross-cutting challenges of computational support for SUDs.

## 5.2 Interdisciplinary Bridges

Based on the discussions of the workshop and the challenges and opportunities identified in this report, we amplify and broaden the call for interdisciplinary bridges noted in prior CCC workshops on health topics noted in prior health-related CCC workshops [38,53]. Given the diversity of communities affected by SUDs, the diversity of approaches to SUD prevention and treatment, and the diversity of individual experience in recovery, effective computational support can best be designed, implemented, and understood within diverse interdisciplinary teams. Current approaches are frequently siloed by the practices and values of a particular discipline or practice — for example, the focus on novelty as an intellectual merit criterion in NSF-supported computing work versus focus on establishing efficacy in NIH-supported clinical trials. Additionally, fields that may be able to make significant contributions to the design of engaging novel interventions (i.e., the arts) are frequently excluded from the conversation all-together. When interdisciplinary collaborations do occur, they frequently position one discipline in service of another — for example, a technologist collaborating with a clinician to recruit participants to use some novel technology or a clinical researcher hiring a technologist to implement an already-conceptualized digital intervention. These approaches do not allow either side to fully benefit from the expertise of the other and likely lead to dissemination which is also siloed in specific communities of practice. There have been successful models for supporting truly interdisciplinary work in other areas. One example may be in the funding of big interdisciplinary centers of excellence, such as the NIH-funded MD2K[4] ("Center for Excellence for Mobile Sensor Data-to-Knowledge") which seeks to develop tools to gather, analyze, and interpret health data generated by mobile and wearable sensors and includes participants from over a hundred institutions. A smaller-scope example may include the annual Jacobs Foundation Annual Conference[5], which brings together junior researchers from different disciplines doing work on a particular societal challenge and provides them with mentorship and seed funding to support an interdisciplinary research project. Others have also pointed to the importance of operational support for this kind of bridge building, such as creating and utilizing shared infrastructures for data collection and donation efforts [12]. It is clear that for truly interdisciplinary work to be possible in the context of computational support for SUDs, it will require substantial support and investments at operational, relational, and funding levels.

---

[4] https://md2k.org/
[5] https://jacobsfoundation.org/en/activity/jacobs-foundation-conference/

## 7. Workshop Attendees

| First Name | Last Name | Affiliation |
|---|---|---|
| Konstantin | Aal | University of Siegen, Germany |
| Alexander | Adams | Cornell University |
| Suzanne | Bakken | Columbia University |
| Olivier | Bodenreider | NIH |
| Alan | Borning | University of Washington |
| Jonathan | Braswell | Basic Meeting List Toolbox |
| Patricia | Cavazos | Washington University |
| Stevie | Chancellor | Northwestern University |
| Yunan | Chen | University of California, Irvine |
| Munmun | De Choudhury | Georgia Institute of Technology |
| Khari | Douglas | CRA/CCC |
| Daniel | Epstein | University of California, Irvine |
| Danny | Gershman | Basic Meeting List Toolbox |
| Kate | Gliske | Hazelden Betty Ford Graduate School of Addiction Studies |
| Peter | Harsha | CRA |
| Nicholas | Jacobson | Dartmouth University |
| Niranjan | Karnik | Rush University |
| Daniel | Keefe | University of Minnesota |
| Cliff | Lampe | University of Michigan |
| Sarah | Lord | Dartmouth University |
| Gabriela | Marcu | University of Michigan |
| Rajalakshmi | Nandakumar | Cornell University |
| George | Nitzburg | Columbia University |

*The participant list continues on the next page.*

| First Name | Last Name | Affiliation |
|---|---|---|
| Srinivasan | Parthasarathy | Ohio State |
| Jessica | Pater | Parkview Health |
| Chanda | Phelan | University of Michigan |
| Timothy | Piehler | University of Minnesota |
| Mashfiqui | Rabbi | Harvard University |
| Thomas | Reese | University of Utah |
| Stephen | Schueller | University of California, Irvine |
| Ann | Schwartz | CRA/CCC |
| Katie | Siek | Indiana University |
| Margorie | Skubic | University of Missouri |
| Laura | Starbird | Columbia University |
| Jacob | Sunshine | University of Washington |
| Ronald | Tannenbaum | InTheRooms |
| Sebastian | Taugerbeck | University of Siegen, Germany |
| Brooks | Tiffany | Kaiser Permanente |
| Daniela | Tudor | WeConnect Recovery |
| Tiffany | Veinot | University of Michigan |
| Tisha | Wiley | NIH |
| Helen | Wright | CRA/CCC |
| Goli | Yamini | NSF |
| Chris | Yang | Drexel University |
| Lana | Yarosh | University of Minnesota |
| Chuang-Wen | You | National Taiwan University |

**NOTES**

**NOTES**

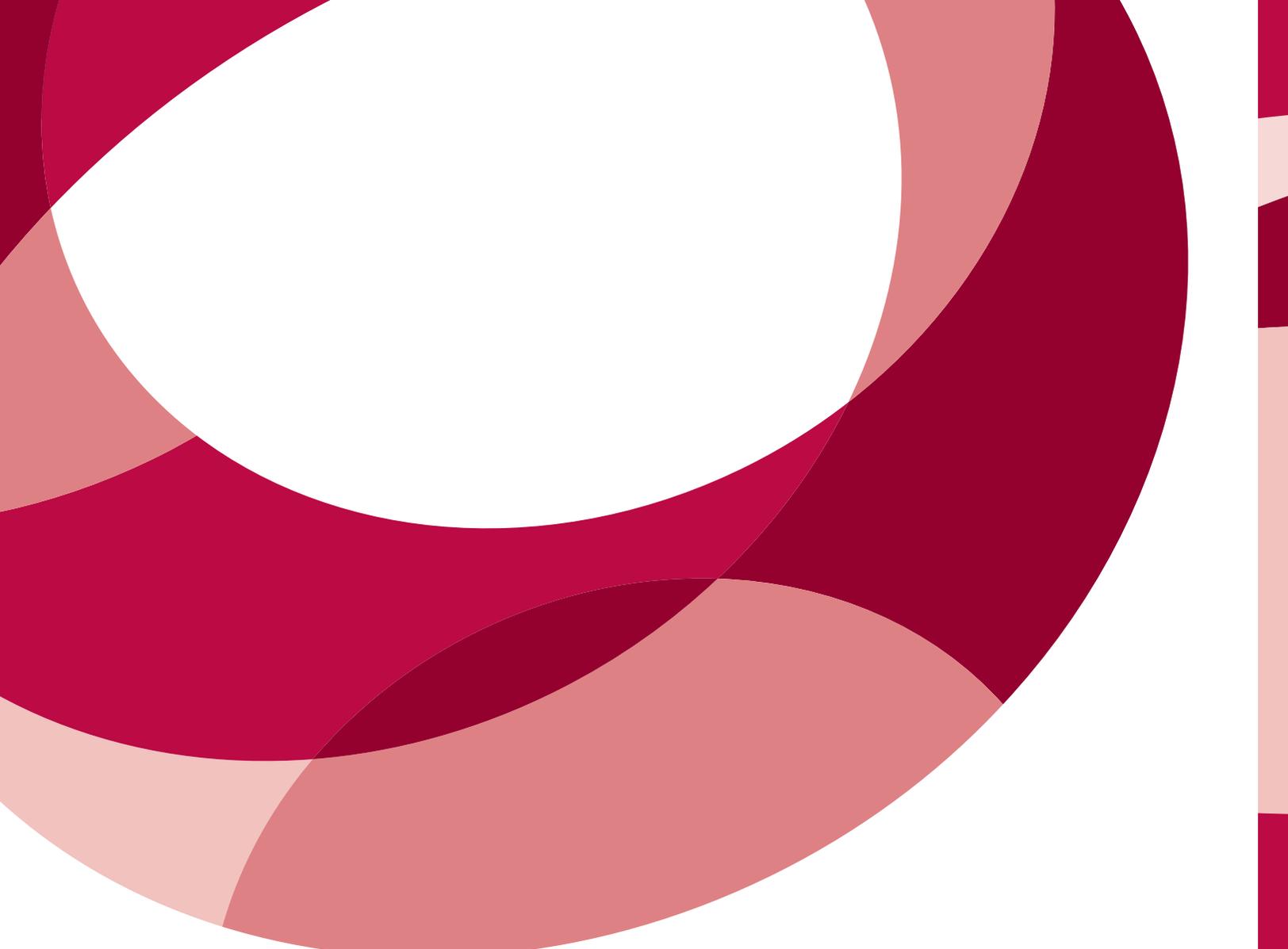

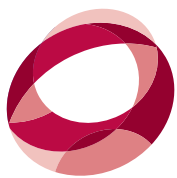

**CCC**
Computing Community Consortium
Catalyst

1828 L Street, NW, Suite 800
Washington, DC 20036
P: 202 234 2111 F: 202 667 1066
www.cra.org cccinfo@cra.org